\documentclass{article}

\usepackage{amsfonts, amsmath, amssymb, amsbsy}
\usepackage{epsfig}
\usepackage{graphicx}
\usepackage{longtable}

\begin{document}

\begin{center}
\textbf{Neutrinos: The Big Question and Physics Opportunities}
\end{center}

\begin{center}
\textbf{Thomas Strauss$^1$}
\end{center}

\begin{center}
{\it
\noindent $^1$University of Bern\\
Albert Einstein Center for Fundamental Physics for Bern\\
Laboratory for High Energy Physics\\
Sidlerstrasse 5, CH-3012 Bern, Switzerland
}
\end{center}

\begin{abstract}
This article summarises a talk given at the 2014 Palermo workshop on Astrophysics. It covers a short review on the neutrino physics status and the potential physics opportunities of future experiments. During the last year our knowledge on the neutrino oscillation parameter $\sin^2\theta_{13}$  improved dramatically, and the large value opened the way to oscillation experiments sensitive to possible CP-violation. The first high-energetic neutrinos in the TeV range were detected in the IceCube experiment, while the Planck collaboration set further limits on the number of active neutrinos from cosmological constraints. Over the next years the Katrin will investigate the beta decay of Tritium to study the absolute neutrino mass scale, while new experiments will investigate the potential sterile neutrino scenario which could explain the event excess of the MiniBooNE and LSND experiment.

\end{abstract}

\section{Introduction}
Neutrinos are the lightest known particles with less than 1/4'000'000 mass of an electron. They have no electrical charge, thus their interaction with other particles is limited to the weak force and on a much smaller scale gravitation. We have discovered three distinct neutrinos, which are named after the partners we observe them with in charge current interactions, the electron-, muon and tau-neutrino. They are the second most numerous particle in the universe, and are most remarkable.  Even though they are very light, the total amount of 330 neutrinos per cubic centimetre results in a measurable influence of neutrinos on the clustering of galaxies. 

As only weakly interacting particles, the cross section of neutrinos with other particles is very low. Yet the rates of neutrino production are large. The sun neutrino flux results in 10$^{11}$ neutrinos per square centimetre and second at a distance of about 1.5 million kilometre; a nuclear reactor of 100\,MW will produce a rate of 10$^{13}$ neutrinos per square centimetre and second at a distance of 10\,m, for comparison. 

As presented in \cite{RevModPhys841307}, our understanding of neutrinos is driven by their interactions with matter. Depending on the energy of the neutrino, one can classify their origin. Neutrinos are messengers over a large band of energies. From 10$^{-2}$\,eV on we attribute them as coming from the BigBang, while terrestrial neutrinos are becoming present at energies 1\,eV. Slightly more energetic are neutrinos produced in reactors by nuclear fission or fusion. Higher energies are reached when neutrinos are produced in galactic events like supernovae with energies up to a few 10$^{4}$\,eV. Atmospheric neutrinos are produced by the decay of secondary particles produced in the earth's atmosphere and can reach up to hundreds of TeV. Similarly particle accelerators can produce neutrinos, usually within the GeV range. Neutrinos from active galactic nuclei can have energies up to the EeV range. Above that the neutrinos are assumed to originate from proton interactions with the cosmic microwave background. 

While there are many sources, not all are accessible with today's detectors. The highly energetic neutrinos require massive detectors; the biggest currently available is the IceCube experiment in Antarctica with a volume of 1 cubic-kilometre of instrumented ice, sensitive to the PeV range. For the lower energies the limits for detection are given by experiments sensitive to rare events, which are reaching down to a few electron volt and are sensitive for terrestrial neutrinos. 

The large complication in neutrino physics is the observation of them via their matter interaction. Contrary to charged particles, neutrinos are only observed in their 'destruction' or 'creation'. The measurement is always dependent on the neutrino flux, the cross section, nuclear effects and detector efficiencies. The weak force interaction is analytical solvable only for hydrogen; all other targets are too complicated due to the multiple protons and neutrons in the nuclei. Either charged current or neutral current are mediated by vector bosons, and a parametrisation of the target nucleus is possible. Yet, the main problem is the possible interaction of the produced particles within the nuclei before they can be detected, the finite state interactions. Its description is challenging as the initial state and what escapes the nuclei can be totally different. Theorists are using nuclear physics approaches to shed light on this interesting area of neutrino physics.

\section{Geo neutrinos}
\begin{figure*}[!ht] 
\begin{center}
\includegraphics[width=120mm,height=80mm,angle=0] {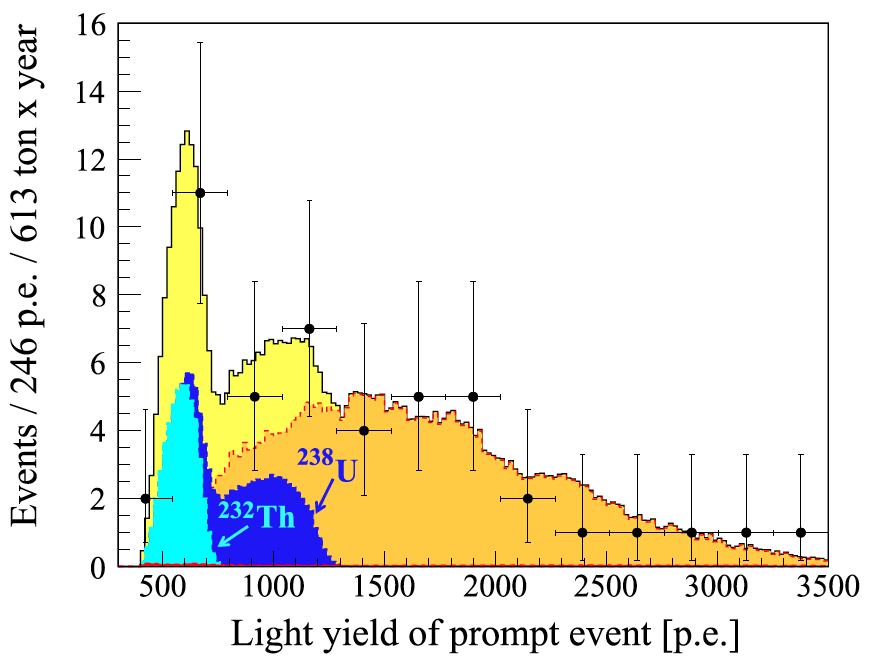}
\caption{Prompt light yield spectrum of Borexino. The 46 anti-neutrino candidates
and the best fit with free U (blue) and Th (cyan) contributions. The yellow
area isolates the total contribution of geo-$\bar{\nu}_{e}$'s. Dashed red line/orange area shows the contribution from reactor-$\bar{\nu}_{e}$'s. The conversion from p.e. to energy is approximately 500 p.e./MeV. Figure is taken from \cite{borexino}.} \label{fig:borexino}
\end{center}
\end{figure*}

Neutrinos from geological source are due to the decay of unstable isotopes in the earth's crust and mantle. Their main contribution is from Thorium and Uranium decays, naturally present in the mantle. This source of low energetic anti-neutrinos was recently detected by experiments with a very high sensitivity for rate, low energetic events, Borexino \cite{borexino} and KamLand \cite{kamland}. They provide not only an understanding of the neutrinos, but also the chance to study the heat output of the earth. Shown in Figure \ref{fig:borexino} is the result of the Borexino experiment, their observed energy spectra for $\bar{\nu}_{e}$ events, with 46 anti-neutrino candidates. Their best fit uses free Uranium and Thorium contributions. In yellow they sum the total contribution of geo-$\bar{\nu}_{e}$'s. The dashed red line/orange area shows the contribution from reactor-$\bar{\nu}_{e}$'s. Their background in red is negligible, and the energy range of the plot reaches from about 1 to 3.5\,MeV. This novel result, similar to the one from the KamLand experiment gives the first glimpse into the earth.

\section{Galactic neutrinos}
Neutrinos from non terrestrial sources are either solar, atmospheric, cosmic or supernova neutrinos. They are unique messengers of far-away phenomena. They do not get deflected by interstellar magnetic fields. Thus, with a detector with a good pointing resolution one might use the events to point back to their sources. Neutrinos rarely interact with matter, except for the Glashow resonance, which is a resonance from the formation of an intermediate W-boson. Neutrinos traverse regions where no light can come from, like nebula or even the inside of our sun. The neutrinos produced inside will reach earth within 8\,min, while the light from the same fusion process will need more than 100'000 light years to escape the core of the sun. Thus, the neutrinos carry information about events with the highest energies and the most distant phenomena, allowing 'neutrino astronomy'. The cut-off at high energies, the GreisenÐZatsepinÐKuzmin \cite{gzk} limit is a theoretical upper limit of the energy from particles from distant sources at 5$^{19}$\,eV, induced by the energy loss of cosmic ray protons with the microwave background radiation over large distances. Here neutrinos offer the potential to see new physics.

When the universe was created in the Big Bang, the leftover photons have an energy of $2.72548\pm0.00057$\,K, creating the cosmic microwave background - a nearly perfect black body. While there are many other interesting topics associated with the Big Bang, like the origin of the baryon asymmetry or the origin and type of dark matter and dark energy, neutrinos were also present. They froze out at a temperature of about 2.5\,MeV, well before the electron-positron annihilation period took place. Under this assumption, we can expect leftover neutrinos related to the Big Bang with an energy of T=1.95\,K or E=170\,$\mu$eV. These small energies make a detection unlikely, due to the requirements on the energy resolution of the detector and low thresholds with a large background. Yet, without neutrinos, the cosmic microwave background cannot be explained. This knowledge was used by the Planck experiment \cite{planck} to calculate the number of effective neutrino masses from their measurement. Alone they measured 4.53$^{+1.5}_{-1.4}$, combined with other experiments and theoretical predictions this number converges to 3.28$^{+0.67}_{-0.64}$.

\begin{figure*}[!ht] 
\begin{center}
\includegraphics[width=130mm,height=60mm,angle=0] {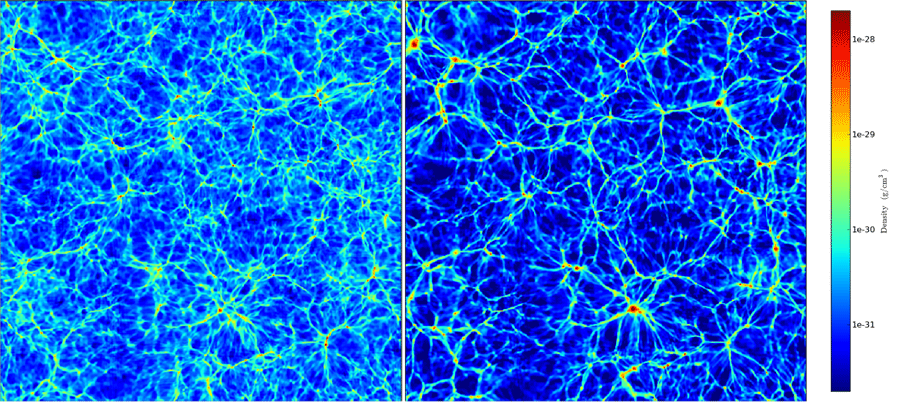}
\caption{Comparison of density distribution in the Universe with (left) and without (right) massive neutrinos. The maps are based on numerical simulations, the colours account for the density of ordinary (baryonic) matter in one slice of the simulation box. The two simulations started from the same initial conditions, with either M$_\nu$=0 (right) or M$_\nu$=1.9\,eV (left). In the massive neutrino case, matter is spread over a larger number of structures and there is less density contrast. (The unrealistically large neutrino mass of 1.9\,eV was chosen so as to make the comparison clear.), Figure is taken from \cite{clustering2}.} \label{fig:clustering}
\end{center}
\end{figure*}

Another effect of the neutrinos on galactic scales is due to their tiny, but non-zero mass. The sheer amount of neutrinos results in them having influence on the clustering of galaxies. From oscillation measurements we know their summed mass to be larger than 0.05\,eV, and the lower limit from direct mass measurements place their sum below 2\,eV. Cosmologically the sum should be less than 0.3\,eV, as explained in \cite{clustering}. Shown in Figure \ref{fig:clustering} is the mass effect on the density distribution in the Universe. Both maps show numerical simulations starting from the same initial conditions, one for  M$_\nu$=0 (right), the other for M$_\nu$=1.9\,eV (left). For massive neutrinos, the matter is spreading over a larger number of structures and with less density contrast. For illustration an unrealistic high neutrino mass was chosen in these images. The remaining question that cannot be answered is the contribution of the individual neutrinos to the mass sum. 

Supernova type II, core collapse supernova, are a powerful source of neutrinos. Nearly the whole gravitational energy of the collapse is radiated away by neutrinos (99\,\%), and these explosions are among the highest recorded energetic processes in the universe, with energies of up to 10$^{53}$\,erg \cite{supernova}. For these energies, all species of neutrinos are produced in electron-positron annihilation processes $e^+e_-\rightarrow\nu_i\bar{\nu}_i$. Only the neutrinos allow insight to the inner workings of the supernova. The neutrinos are emitted within a burst of a few tens of seconds duration, ranging energies of a few tens of MeV. The light emitted form the collapse will emerge later. Thus, neutrinos can be used to trigger the observation with optical telescopes. A network of neutrino experiments \cite{snews} is constantly watching for events. So far only the Supernova 1987A was recorded \cite{1987}. The one and only one such event released more neutrinos than our sun will produce in its life. The neutrinos arrived a few hours before the light. In total we have seen 24 neutrinos from three different experiments. All experiments were only sensitive to $\bar{\nu}_e$. Future experiments in preparation aim to be able to detect the neutrino component too. The rate of the Supernova Type II collapse in the near vicinity (meaning that today's experiments are able to detect an increase in the neutrino event rate) is not precisely known, but from historical source one can estimate them to be in the range of 30-50 years, which means that we can expect another one sooner or later.

The main drinking question of neutrino and astrophysicists related to Supernova neutrinos are:
\begin{itemize}
 \item How do core-collapse supernovae explode?
 \item How do they form neutron stars and black holes?
 \item What are the nucleo-synthesis products of supernovae?
 \item What are the actions and properties of neutrinos?
 \item What is the cosmic rate of black hole formation?
 \item Which supernova-like events make neutrinos?
 \item What else is out there that makes neutrinos?
 \item What is the time profile and energy spectra of the neutrinos?
 \item Can we measure the relic supernova neutrinos?
\end{itemize}

Only upon observation of a type II supernova with the next generation of experiments are we able to answer these questions.

Astrophysical sources of neutrinos emit at energies larger than TeV scale. These neutrinos are a by-product of cosmic ray - matter collisions. The interaction of the protons with the cosmic ray background creates the GZK neutrinos. Other high energy neutrinos are produced in active galactic nuclei (AGN) and gamma ray bursts (GRB). Another possible source could be Weakly Interacting Particles (WIMPs). The only experiment currently able to measure neutrinos in the PeV range is the IceCube detector, situated at the South Pole, 1.5\,km below the surface in the arctic ice shield. The detector uses the ice shield as target medium. 5160 PMT's are mounted on 86 strings and lowered into the Glacier, up to 3\,km below the surface. The instrumented detector has a volume larger than 1 cubic kilometre. To date the experiment has released data about 23 high energetic events with more than 100 TeV, extending into the PeV range, with the first of the events named after Sesame Street characters 'Ernie' and 'Bert' \cite{icecube}. Naturally there is a large interest in these events, as they might point to high energetic sources in the sky, straight back to the source. For the very energetic events, these sources are likely extra-galactic. Due to the low statistics and pointing uncertainty no source in the sky has been identified yet. Further detector upgrades aim to increase the detectors capabilities.

Ultra high energy neutrinos are interesting because they will answer questions as basic as:
\begin{itemize}
\item Do ultra high energy neutrinos exist?
\item At which energies do they exists?
\item What shape is the energy distribution?
\item Can we detect GZK neutrinos of EeV?
\item Do even higher energetic neutrinos then GZK neutrinos exists, and where would they come from?
\end{itemize}

\section{Neutrino oscillation}
\begin{figure*}[!ht] 
\begin{center}
\includegraphics[width=120mm,height=80mm,angle=0] {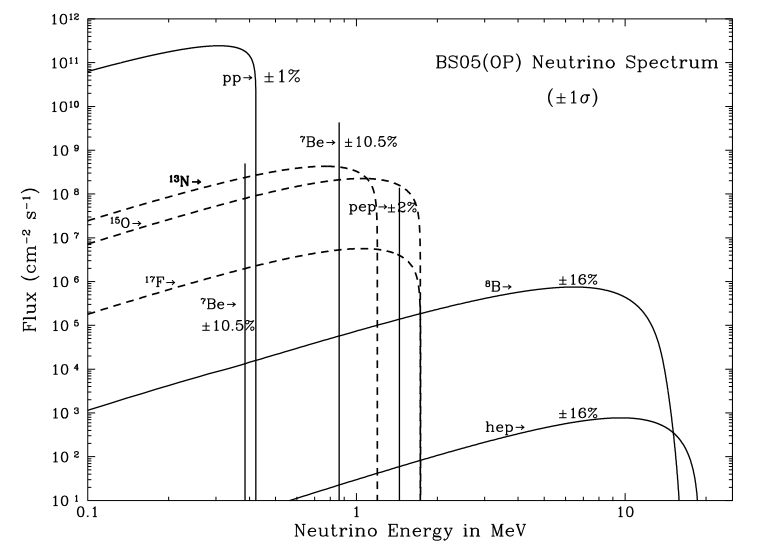}
\caption{Solar neutrino energy spectrum for the solar model BS05(OP). The uncertainties
are taken from Table 8 of Bahcall and Serenelli (2005). Figure is taken from \cite{bahcall}.} \label{fig:bahcall}
\end{center}
\end{figure*}

The solar neutrino flux on earth was first calculated by J Bahcall in the 1960's. The nuclear processes in the sun's core produces a huge number of 10$^{38}$ neutrinos per second via the proton-proton fusion chain. Only electron-neutrinos are produced in the sun, with an energy spectra below 20\,MeV, as shown in Figure \ref{fig:bahcall}.

Solar neutrinos gave the first surprise of neutrino physics, when R. Davis jr. famous Homestake mine experiment started data taking. The experiment used an underground tank of 8.5\,m diameter and 16\,m length surrounded in a water shielding filled with  400Õ000\,l Perchloro-ethylene, C$_2$Cl$_4$ (or Dry-cleaner), to capture neutrinos via the inverse beta decay in the reaction $^{37}Cl + \nu\rightarrow ^{37}\,Ar + e^+$. The energy threshold of E$_\nu>$\,0.8\,MeV made the experiments sensitive to the $^8$B fusion process only. The produced $^{37}$Ar has a decay half-life of 35 days. By counting the argon atoms, the rate of neutrinos could be extracted, and was constantly only one third of the predicted flux. The repurposed KamioKaNDE-II detector filled with 2000\,t of water was able to measure the directionality of the solar neutrino events via the Cerenkow light of particles produced in neutrino interactions, and was also off in the event count by about a factor 2. Finally in 2001 the SNO experiment was able to compete the picture by measuring not only the flux of electron neutrinos, but of all neutrinos types. They used a heavy water target which is sensitive to all neutrino flavours in their neutral current interactions. They could filter the quasi-elastic scattering of electron events and the charged current interactions, and combine the total flux. The combined rates showed an agreement with the prediction of Bahcall's solar model. In the emerging picture one started to understand that neutrinos changed flavour while traveling, a prediction already made in 1970's by Bruno Pontecorvo. This revelation implied that neutrinos have a mass; yet, it is very small as they are traveling nearly with the speed of light.

Neutrinos in the atmosphere are produced by the decay of secondary particles from initial cosmic ray particles. Protons or Helium atoms smashing into earth's atmosphere create an air shower, in which pions, muons and electrons are produced. The resulting decay of pions to muons and the muons to electrons creates a large flux of neutrinos, with a ratio of two to one for the flavour between $\nu_\mu$ and $\nu_e$ (due to the $\pi/\mu$ cascade. Their energy range is in the GeV's, as opposed to the solar MeV range. The highest energies for atmospheric neutrinos have a cut-off at the 100TeV scale. As neutrinos travel through matter unhindered, underground detectors are sensitive to neutrinos from all possible directions. The Super-KamioKaNDE experiment in Japan was the first to unambiguously detect the neutrino oscillation with atmospheric neutrinos in 1998. They not only could extract the missing azimuthal shape of the neutrino showing a lack of neutrinos for particles traveling through earth, but also an energy spectra scaling with the L/E ratio, the oscillation frequency. The path length for atmospheric neutrinos ranges from 10 to 13'000\,km. 

Today's picture of neutrino oscillation is described by the PontecorvoÐMakiÐNakagawaÐSakata matrix. Depending on the representation one can split this matrix in sub-matrices, one explaining the atmospheric oscillation, one for the solar puzzle and one for the remainder. This splitting is due to the different oscillation frequency between solar and atmospheric neutrinos. The oscillation scheme is modified by matter effects for electron neutrinos, the MikheyevÐSmirnovÐWolfenstein effect, as, due to the abundance of electrons in matter, the energy levels of the neutrino mass eigenstates for electron neutrinos is changed as a result of charged current coherent forward scattering. In 2013 the Daya Bay \cite{dayabay} and T2K \cite{t2k} collaboration published their findings on the last unknown mixing angle $\theta_{13}$ and its rather large value allows future experiments to determine the neutrino mass hierarchy and the possible exploration of CP violating effects in the neutrino oscillation. For the future large underground detectors have been proposed to investigate both mass hierarchy and the phase of the possible CP violation. The current design discussion places a multi kiloton detector in a new neutrino beam line from Fermilab to Homestake mine \cite{lbne}. The goal is to observe the neutrinos oscillations, and investigate spectral distortions of the oscillation frequency, which will be enhanced or reduced depending on the mass hierarchy and CP phase. Furthermore a comparison of neutrino and anti-neutrino events will allow a direct evidence of a CP violating effect. The project's goal and timeframe is based on a 20+ year schematic. The detector of choice is a liquid argon time projection chamber. The main reason for the liquid argon technique instead of the water cerenkov approach used for very large detectors is the improvement in imaging capabilities of the neutrino interaction, especially for the investigation of final state interactions, and the ability to distinguish a shower induced by photons from showers created by electrons. 

Due to the observation, we know that neutrinos must have small mass difference, with a total mass non-equal to zero. One experiment in preparation, the Katrin experiment \cite{katrin}, aims to measure the end point of the beta decay spectra of Tritium. With a good enough energy resolution the rates should be affected from the mass, which needs to be subtracted from the energy available in the decay. The spectrometer of the experiment is as large as a four story building, with one of the most sensitive magnetic systems world-wide. It is in the construction phase and will require lots of statistics to prove their energy resolution and any results. The mass generation of neutrinos can be described in different ways; the most prominent theories are the the neutrino is Majorana, e.g. its possible to be its own anti-particle. Experiments investigating the neutrino-less double-beta decay are careful studies of nearly background free experiments based on the few isotopes that actual have a double beta-decay. They are analysing this spectra to observe a peak which will be at the maximum for the summed energy of the emitted electrons, compared to the case where neutrinos are emitted and take away part of the energy. Majorana neutrinos could be used to fit into the theory of leptogenesis. The other possible explanation for neutrino mass is that it is a Dirac particle. This allows us to introduce a right handed, sterile neutrino, which could be a candidate for Dark Matter.

The questions currently driving the field are:
\begin{itemize}
 \item How do neutrinos gain mass?
 \item What is the ordering of the neutrino mass eigenstates?
 \item Is the neutrino its own anti-particle?
 \item Can we observe CP violation in the leptonic sector?
 \item If neutrino are Majorna particles, what are the 2 Majorana phases?
 \item Why are neutrinos light and neutral?
 \item Why are the neutrino mixing parameters so unlike to the quark mixing?
 \item What is the Octant of $\theta_{23}$?
 \item Are neutrinos always left-handed, and anti-neutrinos right handed, or not?
\end{itemize}

\section{Exotic Neutrino Physics}
The measurement of the Z-boson decay width by the LEP experiments gave the number of active neutrinos to be three. Yet it did not exclude the existence of possible sterile neutrinos, not weakly interacting\cite{lightsterile}. The idea is based on the signal excess observed by the LSND and MiniBooNE experiments on short baseline neutrino beams \cite{conrad}. The LSND experiment saw a oscillation for a L/E ratio of 1\,m per MeV, which is in disagreement to the standard 3 flavour oscillation. The MiniBooNE excess is a different electro-magnetic excess at lower energies. Further hints are based on the re-analysis of the neutrino events from reactor based experiments \cite{reactor,reactor2,reactor3}. After the re-calculation for the flux, there is a discrepancy for very short distances with missing events, that is not ruled out by cosmic constraints. Other hints are based on the missing events from Gallium source experiments. While there are hints of a non-oscillation nature of these excess, the theoretical explanation of a fourth (or more) sterile neutrino sounds appealing \cite{giunti}, and will be investigated in the near future by several new experiments \cite{microboone,lar1nd}. The appeal comes from the easy fit of a heavy sterile neutrino with the SeeSaw mechanism. The questions here are:
\begin{itemize}
 \item Are there sterile neutrinos?
\item Why would they exist?
\item Why would they be light or heavy?
\item What is their mass?
\item What can we say about them?
\item Are they candidates for Dark Matter?
\item Is there one, or more?
\end{itemize}

There are further results in neutrino physics that cannot be explained within the Standard Model. One of them is the electron capture decay of hydrogen like atoms like $^{142}$Pm, $^{122}$I and $^{140}$Pr at the storage ring at GSI \cite{litinov}. There the time decay of the atoms seems to be super-imposed with a 7\,s oscillation, instead of the expected simple exponential decay. The rates of production of different final states contribute to the total event rate incoherently. There are (at least) 3 neutrino mass eigenstates $\nu_i$ with unequal masses $m_i$. For a hydrogen like beta-decay $^{142}$Pm$\rightarrow^{142}$Nd + $\nu$ in which a parent particle P decays to a daughter particle D plus a neutrino, there are 3 distinct final mass states $\nu_i$:
\begin{itemize}
\item P$\rightarrow$ D + $\nu_1$
\item P$\rightarrow$ D + $\nu_2$
\item P$\rightarrow$ D + $\nu_3$
\end{itemize}
The decay rate is the sum of all the 3 processes: $\sum\limits_{i=1}\left\{\frac{dN}{dt}(^{142}Pm \rightarrow ^{142}Nd + \nu_i; t)\right\}$, and unlike to the flavor neutrino oscillation, this sum is not expected to depend on the mass splitting. This dependence comes from interference. 

Theorists are trying to be bold and question the established theorem of CPT conservation \cite{kayser}. The neutrino oscillation would be different for neutrinos and anti-neutrinos, so oscillation experiments are providing a stringent probe in the CPT conservation. Furthermore, the neutrino-less beta-decay would show a lepton-number violation and Majorana mass. While the theories sound wild, these 'Gedankenexperiments' show the need for more precise measurements in the neutrino oscillation sector, where the oscillation parameters are not precisely determined yet. The goal of statistical and systematic errors in the few percent range seems plausible over the next generation of experiments.

\section{Conclusion}
Neutrinos are special, not only because they have a small mass and are elusive, but because they have the weakest interactions within the Standard Model, so they are most sensitive to non-SM interactions. They are the lightest known particles by far, and we still have no full understanding why. Their flavour oscillation is sensitive to the very tiniest effects, opening the field for discoveries of new physics. They are the only electrically neutral fermionic constituent of matter, which is puzzling. They are the only known candidate for Majorana mass, which is non-Standard Model. So far we have only observed left handed neutrinos, while anti-neutrinos always right handed. Due to their low cross section a direct measurement of this is nearly impossible. While there might be a CP violation for the oscillation appearance probability, CPT requires the disappearance of neutrino and anti-neutrino to be equal. Any divergence would be stunning indeed.

Neutrinos are messengers over the largest energy ranges, from different sources and times. They excite us because we are surprised by nearly every measurement. While we believe we know some things about them we still believe that we have not learned all we can about them. They allow us to probe for non-SM physics in the lepton sector, and over the next decades we have interesting results ahead of us.

\bigskip
\bigskip
\noindent {\bf Acknowledgments}\,\,\, I thank the Albert Einstein Center for fundamental particle physics for Bern at the University of Bern for the kind financial support for traveling to the conference. I like to thank the workshop organisers for the invitation and nice days at the conference, and their support for the infrastructure and arrangements. The talk at the conference and the proceedings profited heavily from the Fermilab lecture series: ÔThe allure of ultrasensitive experimentsÕ.

\end{document}